\documentclass{article}
\usepackage{graphicx} 
\usepackage[left=1in, right=1in, top=1in, bottom=1in]{geometry}
\usepackage{amsmath}
\usepackage[dvipsnames]{xcolor}
\usepackage{amssymb}
\usepackage[sort, numbers]{natbib}

\usepackage{authblk}

\title{Support for Gravitationally-Attractive Composite Antimatter and Gravitationally-Repulsive Non-composite Antimatter}
\author[1,2,3]{Giovanni Maria Piacentino}
\author[4]{Anthony Palladino}
\affil[1]{UniNettuno University, Rome, Italy}
\affil[2]{INFN, Sezione di Roma Tor Vergata, Rome, Italy}
\affil[3]{INAF, Osservatorio Astronomico di Roma, Monteporzio Catone, Italy}
\affil[4]{Boston University, Boston, USA}
\date{28 March 2025}
\newcommand\blfootnote[1]{%
  \begingroup
  \renewcommand\thefootnote{}\footnote{#1}%
  \addtocounter{footnote}{-1}%
  \endgroup
}

\begin{document}

\maketitle

\begin{abstract}

The ALPHA collaboration recently published the article ``\emph{Observation of the
effects of gravity on the motion of antimatter}'' in issue 621 of Nature (2023), in which
their conclusion states that ``\emph{The probability that our data are consistent with 
repulsive gravity is so small as to be quantitatively meaningless (less than $\it{10}^{-\it{15}}$). Consequently, we can rule out the existence of repulsive gravity of magnitude 1$g$ between the Earth and antimatter.}'' Indeed, their experiment proved that antihydrogen falls toward the Earth. Antihydrogen, however, is a composite particle and its mass, like that of the proton, is dominated by the mass of its binding energy. In this paper, we point out that their experiment only measures the gravitational effect on \emph{composite} antimatter, and not free or valence antimatter particles such as antiquarks or antileptons.
Here, we review the various theoretical frameworks supporting the possibility that valence antiquarks experience a repulsion of the order $g$ by the Earth's gravitational field. We show that the ALPHA collaboration's recent result may actually be consistent with a repulsive gravitational component of about 15\%. Next, we review our concept for a direct interaction
experiment between antiquarks and the Earth's gravitational field, which
should be capable of providing irrefutable evidence of the gravitational repulsion of non-composite antiparticles. Finally, we suggest that existing data, collected by AMS-01 of daughter particles from cosmic protons interacting with the Mir space station, may help demonstrate the feasibility of our proposed experiment. 
\blfootnote{A condensed version of this manuscript to appear in the Proceedings of the 59$^{\mathrm{th}}$ Rencontres de Moriond Conference on Gravitation, 2025.}
\end{abstract}

\section{Mass decomposition}

It is well known that a gravitational system composed of two masses orbits, in first approximation, around a massive body as if its mass had a correction due to the mass of the binding potential energy. This problem was well addressed and resolved in the case of charged interacting spheres considering intrinsic and electromagnetic mass by Brillouin in ``Relativity Reexamined'' in 1970 \cite{brillouin-1970}. Systems composed and held together by fields are not only characteristic of astronomical structures such as the Earth and Moon, or Jupiter and its satellites, but also of mesons and baryons---composite systems of quarks bound together by the gluonic field such that their gravitational mass is composed of that of the valence quarks plus that of the binding energy. In the case of protons this decomposition is not exhaustive because the valence quarks are continuously in motion and it is necessary, in principle, to add both vibrational and rotational (spin dependent) kinetic energy components, giving rise to a pseudo-Jacobi integral of the internal motion of the composite particle. 
Thus, the total mass of the proton, a composite particle, can be expressed as:
\begin{equation}
M_{\mathrm{p}}=2m_{\mathrm{u}}+m_{\mathrm{d}}+(U_{\mathrm{binding}}+U_{\mathrm{kinetic}}+U_{\mathrm{spin}})/c^{2} \,\,.
\end{equation}
This distinction on the origin of the proton mass and, more generally, on that of all composite particles, must also apply to antibaryons, antimesons, antiatoms, etc. 

\section{Gravitational interaction of antiparticles}

There are at least three theoretical frameworks for describing the gravitational effect on non-composite antiparticles as being repulsive. We provide a high level summary of these frameworks in the following three subsections. 

 
\subsection{Worldine curvature} 

The Feynman-Stueckelberg interpretation of antimatter states that one can apply time reversal and charge conjugation to matter to produce antimatter. But what kind of time reversal should one use in General Relativity, i.e., in the presence of Gravity? Considerations in the context of a non-quantum-mechanical discussion have been well analyzed in by Recami and Mignani~\cite{mignani_recami}, in which they consider the phenomenology of tachyons. This hypothetical particle is forced to travel perpetually at superluminal speed, seen in a normal inertial reference system it would appear as though it was emitted at the point of decay and absorbed at the point of origin, but as if it was equipped with an opposite charge to the conventional one. According to the principle of ``reinterpretation'' (proposed by Recami and Mignani) they would therefore be antiparticles according to the definition of Feynman-Stueckelberg if observed by a normal, infraluminal inertial reference system. Recami and Mignani did not consider the presence of gravity in their works, but it is intuitive to think that by traveling the same geodetic of a similar subluminal particle but in the opposite sense they must ``fall upwards'' that is, it would appear to be repelled in a gravitational field. Returning to the question of which time should be used for the time reversal in the definition of antimatter, there are doubts that must be its own proper time.

While the Feynman-Stueckelberg interpretation is consistent with classical and quantum electrodynamics, Jos\'{e} Ripalda argued in~\cite{ripalda2010timereversalnegativeenergies} that it is only compatible with General Relativity if the stress-energy tensor of matter going backward in time has a sign that is opposite to the stress-energy tensor of matter going forward in time. This implies that there are two metric tensors for the geometry of spacetime that share the same stress-energy tensor with opposite signs; meaning that they interact through repulsive gravity. In particular, starting from the ``length'' definition  of a particle's worldline  (with $c=1$), 
parameterized by its proper time: 

\begin{equation}
d\tau = \frac{dt^\prime}{\gamma } = \pm \sqrt[]{dx^{\mu }dx_{\mu }}   
\end{equation}
where $\gamma$ can have either sign. Therefore,
\begin{equation}
\begin{split}
m &= m_{0}\gamma \\
u^{\mu } &= \gamma (1,u) \\ 
p^{\mu } &= m_{0}\gamma (1,u)
\end{split}
\end{equation}
And following $\gamma$ also $m$, $u$, and $p$ are odd under proper time inversion, so that
a T transformation $\tau \rightarrow -\tau$ implies that:
\begin{equation}
\begin{split}
m &\rightarrow  -m\\ 
u^{\mu } &\rightarrow -u^{\mu } \\ 
p^{\mu } &\rightarrow -p^{\mu } \\
(x^{1},x^{2},x^{3}) &\rightarrow (x^{1},x^{2},x^{3})
\end{split}
\end{equation}
On the contrary, a PT transformation implies that:
\begin{equation}
\begin{split}
m &\rightarrow -m\\
u^{\mu } &\rightarrow -u^{\mu } \\
p^{\mu } &\rightarrow -p^{\mu }\\
(x^{1},x^{2},x^{3}) &\rightarrow -(x^{1},x^{2},x^{3})
\end{split}
\end{equation}
In fact, the PT reversal of the coordinate system affects all particles and observers, while proper time is a property intrinsic to each worldline, and thus $\tau \rightarrow -\tau$ only affects the worldline in question.  Thus, gravitational interaction between future-pointing particle (non-composite particle) and past-pointing particle (non-composite antiparticle) must be repulsive and the curvatures of spacetime induced by particles and antiparticles are opposite.

\subsection{CPT gravity} 

Massimo Villata suggested different considerations in \cite{villata-2024}: 
``\emph{Combining general relativity and CPT symmetry, the theory of CPT gravity predicts gravitational repulsion between matter and CPT-transformed matter, i.e., antimatter inhabiting an inverted spacetime. Such repulsive gravity turned out to be an excellent candidate for explaining the accelerated expansion of the universe, without the need for dark energy.}'' 
Villata considers CPT transformations in General Relativity, i.e., the simultaneous inversion of the charges and of the direction of time associated with the mirror flipping of all the coordinates of the particles:
\begin{equation}
\begin{split}
\mathrm{CPT:} \,\,\,\, &dx^{\mu }\rightarrow -dx^{\mu }\\
      &q\rightarrow -q\\
      &{} \\
\mathrm{PT:} \,\,\,\, & dx^{\mu }\rightarrow -dx^{\mu }\\ 
\mathrm{C:}  \,\,\,\, & q\rightarrow -q 
\end{split}
\end{equation}
The PT part of the CPT transformation changes the sign of all the components of a four-vector and of a tensor of odd rank but, since in this case it is applied an even number of times, it leaves unchanged the components of a tensor of even rank.  The C part of the transformation means that when the further inversion of the electric charge is introduced, the tensors that contain it change once more in sign and the tensors that are even PT even become odd and vice versa.

Coming now to the geodesic equation, i.e., the equation of motion:
\begin{equation}
\frac{d^{2}x^{\mu }}{d\tau^{2} } =-\Gamma^{\mu }_{\nu \xi } \frac{dx^{\nu }}{d\tau } \frac{dx^{\xi }}{d\tau }    
\end{equation}
In this equation the tensor $\Gamma$ represents the field and the other three elements represent the particle that interacts with it. Each of these elements is odd compared to CPT. If this transformation is applied to all four elements of the equation, we obtain an equation that describes an antiparticle in the gravitational field of antiparticles and the result, similar to the usual one between particles, is that of a gravitational attraction. Conversely, if only the components of the particle are transformed and not those of $\Gamma$, the repulsion between a field generated by ordinary matter and an antiparticle is obtained. Vice versa if only $\Gamma$ is subjected to CPT.

\subsection{Gravitational dipoles} 

A further interesting position is that of Dragan Hajdukovi\'{c}~\cite{hajdukovic-2011-08-DM1, hajdukovic-2012-01, hajdukovic-2012-05, hajdukovic-2014-04, hajdukovic-2016}. He is not so much interested in justifying the repulsion between matter and antimatter as in discovering the prerogatives of empty quantum space. As is well known, quantum mechanical empty space is populated by ephemeral pairs of particles and antiparticles. In the hypothesis supported by Hajdukovi\'{c}, antiparticles have negative gravitational charge, and each particle-antiparticle pair acts as both a gravitational dipole and an electric dipole. The characteristics can be summarized as follows:
\begin{equation}
\begin{split}
\vec{P}_{G} &=\left| m_{0}\right|  \vec{d}  \\
\left| \vec{d} \right|  &\approx \frac{\hslash }{\left| m_{0}\right|  } \frac{1}{c}\\
\left| \vec{P}_{G} \right|  &\approx \frac{\hslash }{c} \,\, .
\end{split}
\end{equation}

If there are $N$ dipoles per unit volume, this preludes to the possibility that, in the presence of an external gravitational field, a gravitational polarization density of the vacuum may arise which can be written as:
\begin{equation}
P_{G_{\mathrm{max}}}=\frac{\hslash A}{c\lambda^{3} }     
\end{equation}
and consequently a mass density given by:
\begin{equation}
\rho_{m_{P_{G}}} =-\nabla \bullet \vec{P}_{G}    
\end{equation}
This could be the origin of Dark Matter, as suggested by Dragan Hajdukovi\'{c}, or with an argument similar to that of Urban et al. \cite{Urban2013TheQV}, it could explain why gravity propagates at the same speed as the electromagnetic field.

\section{Antimatter vs. Antiparticles}

Up to now we have been dealing with the problem of how the presence of antimatter fits into the framework of General Relativity. The problem of what antimatter actually is, however, is much more complex. In fact, in each of the considerations previously illustrated, ultimately belonging to the category of ``matter'' or ``antimatter'' is always linked to the sign (positive or negative) of the proper time of each particle whose value is linked to the ratio $\gamma = E / m_0$. Since $\gamma$ can take both signs, the proper times of the two species of particles evolve intrinsically towards the future and towards the past:
\begin{equation}
d\tau = \frac{dt}{\gamma } = \pm \sqrt[]{dx^{\mu }dx_{\mu }} 
\end{equation}
This type of representation is evidently adherent and satisfactory for point particles, intrinsically endowed with mass (we mean rest mass in the traditional official nomenclature), but what about particles that move at the speed of light, e.g., bosons that transport and mediate interactions? Photons and gluons have no rest mass and consequently their proper time is zero. In their reference system, all events are contemporaneous, i.e., proper time does not exist. A temporal succession of events is introduced by the observer through the filter of the observer's own proper time. The case of massive bosons such as Z, W, and H is different; however, we will not expand on those differences in this paper, as they do not pertain to the present argument. If in our previous description, capable of agreeing with at least the three different theoretical models, presented above, point particles and antiparticles, endowed with intrinsic mass, should present reciprocal repulsive gravitational interaction, the problem of the gravitational interaction of composite particles is more complex. On the other hand, most of the visible mass of the universe seems to be composed of protons and neutrons, composite particles in which most of the mass comes from the binding energy, not from the rest mass of their point components. The mass of this binding energy, in turn, cannot be treated in a simple way because it is composed of a cloud of bosons that mediate the interaction and virtual particles that further complicate the treatment of the problem. 
There are many works on this topic, we will refer to the works of Metz, Pasquini, and Rodini~\cite{pasquini1, pasquini2}, which illustrate a clear overview of the topic and to that of Menary~\cite{menary2024exactlyantimattergravitationallyspeaking}, a member of the ALPHA collaboration.

\subsection{Renormalization of the QCD energy-momentum tensor and mass sum rules}

As commented on previously, protons (antiprotons) are systems composed of quarks (antiquarks) and gluons, and correspondingly, we have for the energy-momentum tensor (EMT) in quantum chromodynamics (QCD):
\begin {equation}
\begin{gathered}T_{}^{\mu \nu}=\  T_{q}^{\mu \nu}+T_{g}^{\mu \nu},\  with\\ T_{q}^{\mu \nu}=\frac{i}{4} \overline{\psi} \  \gamma^{\{ \mu} \overleftrightarrow{D} \ ^{\nu \}} \psi ;\  \  \  \  T_{g}^{\mu \nu}=-F_{}^{\mu \alpha}F_{\alpha}^{\nu}+\frac{g}{4} F_{}^{\alpha \beta}F_{\alpha \beta}.\end{gathered}
\end{equation}
where we mean the sum over the ``flavors'' of the quarks and $D$ represents the covariant derivative. Of course, it is understood that, for both tensors of the second line of the previous equation, the renormalization of the parameters of the QCD Lagrangian is implied. 
Let us therefore recall the state of the art; following \cite{pasquini2} we can consider the decomposition of the proton mass, and implicitly also of the antiproton mass, starting from the elements of the EMT matrix. The general relation is:
\begin {equation}
\begin{gathered}\langle T_{}^{\mu \nu}\rangle \equiv \langle (P^{0},\overrightarrow{P} )|T_{}^{\mu \nu}|(P^{0},\overrightarrow{P} )\rangle =2P^{\mu}P^{\nu};\\ \langle T_{i,R}^{\mu \nu}\rangle =2P^{\mu}P^{\nu}A_{i}(0)+2M^{2}g^{\mu \nu}\overline{C_{i}} (0).\end{gathered}
\end{equation}
where the form factors $A$ and $C$ follow the relations:
\begin {equation}
 A_{q}(0)+A_{g}(0)=1\  \  \  \ \mathrm{and} \  \  \  \  \overline{C_{q}} (0)+\overline{C_{g}} (0)=0,
\end{equation}
and are calculated with zero transferred momentum.
The previous relation implies that both components are related to the proton mass \cite{pasquini2}. Finally, we cite two possible two-term decompositions reported by \cite{pasquini2} and due to \cite{tanaka1, burkert-2023, coriano2025gravitationalformfactorshadrons}. Naturally, decompositions are also possible that take into account contributions in which the decomposition foresees the presence of a non-zero trace term, for example, \cite{tanaka2} obtains a three-term decomposition. In \cite{pasquini1}, they also arrive at producing numerical evaluations of the contributions to the proton mass, and consequently of the antiproton. For a more conscientious discussion we defer to the original references and we are satisfied with only the brief notes and excursions into this delicate topic. In particular, \cite{pasquini1} presents the numerical analysis of the results which, depending on the ``scenario'', contemplates a contribution to the proton mass by the quarks which can reasonably vary from 10 to 20\%, while the contribution from the gluonic field amounts to at least 50\%. Naturally, it is to be assumed that a similar analysis is valid for the antiproton. This suggests that the ALPHA-$g$ measurements can be understood as an experimental verification of these limits. In fact, if we assume that the gluonic part of the contribution to the gravitational mass of the antiproton does not differ from that of the proton~\cite{menary2024exactlyantimattergravitationallyspeaking} and that instead that of the antiquarks, participating in equal percentage to the total, is instead repulsive, the difference in gravitational behavior of hydrogen and antihydrogen should be roughly of the order 10$-$20\%.

\subsection{Consistency with the ALPHA-$g$ experiment's recent result}

\begin{figure}[htp]
    \centering
    \includegraphics[trim={0cm 0cm 0cm 0cm}, clip, width=15.0cm, angle=0]{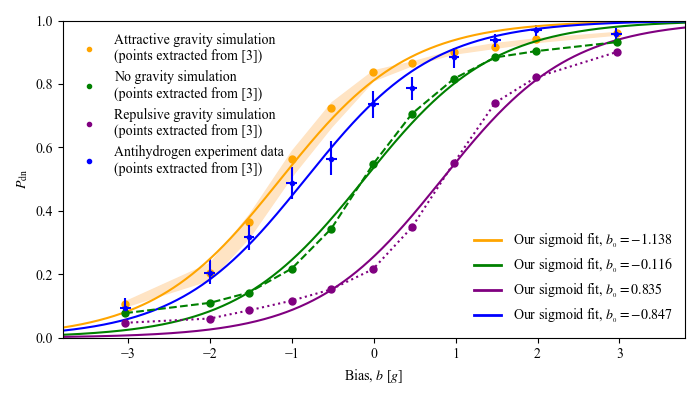}
    \caption{Our sigmoid fits show that the ALPHA collaboration's result may be consistent with a repusive gravitational component of about 15\% (Points extracted from the ALPHA collaboration's paper~\cite{alpha-g-2023}).}
    \label{fig:sigmoid}
\end{figure}

The ALPHA collaboration's recent experiment \cite{alpha-g-2023} measured the gravitational behavior of antihydrogen atoms by letting them migrate and then decay in the Earth's gravitational field with an additional modulatable correction magnetic field. 
We extracted the points published in their figure and reproduced them in our Figure~\ref{fig:sigmoid}. We also fit their measurement and simulations each with its own sigmoid:
\begin{equation}
\begin{split}
P_{\mathrm{dn}} = \frac{1}{1 + e^{-k(b-b_0)}}
\end{split}
\end{equation}
where $b_0$ is the midpoint, corresponding to $P_\mathrm{dn} = 0.5$.

Our fit of their measurement points (blue) corresponding to annihilation events of antihydrogen shows a significant difference from that of the purely attractive simulation (orange). In fact, there appears to be a shift towards the repuslive simulation results (purple). This suggests that the gravitational mass of antihydrogen may be slightly lower than that of hydrogen. How much lower and why? In antihydrogen, there are different valence quarks and the mass of the positron that replaces the electron of the normal hydrogen atom. In the approximation described in Section~1, we consider a gravitational mass connected to the binding energy. While this mass is related to the gluonic field and should not differ between hydrogen and antihydrogen, the mass of the antiquarks could be gravitationally repelled by Earth. This could lead to a difference in total mass equal to at least twice the mass of the valence quarks. If we consider the measurement to be a fractional combination of the attractive and repulsive scenarios, we can parametrize $b_0^\mathrm{Measurement}$ as 
\begin{equation}
\begin{split}
b_{0}^{\mathrm{Measurement}} &= (1-f) \, b_0^{\mathrm{Attractive}} + f \, b_{0}^{\mathrm{Repulsive}} \\
{ } &= (1-f)(-1.138) + f(0.835) \\
\end{split}
\end{equation}
yielding a repulsive component fraction of
\begin{equation}
f = 0.154 \,\, .
\end{equation}
This implies that the ALPHA collaboration's result is consistent with antihydrogen experiencing a repulsive gravitational component of about 15\%. Recall from the theoretical calculations and numerical evaluations in \cite{pasquini1, pasquini2}, we could expect a contribution in the range 10\% to 20\%. 

\section{A new way to weigh antiquarks}

In 1961, Good~\cite{good-1961} calculated using absolute potentials, that a repulsive gravitational interaction of antimatter should introduce a regeneration of neutral  kaons thus resulting in an anomalously large level of CP~violation. At that time, CP~violation in the neutral kaon system was still unknown and the argument appeared strong and elegant, but in light of the measurement of CP~violation~\cite{cronin_fitch-1964} the Good paper could instead turned out to be a very strong argument in favor of antigravity.
In fact, after an initial proposal by Wolfenstein regarding the presence of a superweak force responsible for the violation, in the Standard Model the interaction between quarks and charged gauge bosons has been described by the following term of the Lagrangian:
\begin {equation}
L=\frac{g}{\sqrt{2}} W_{\mu}^{+}\overline{u_{i}} \gamma^{\mu} (1-\gamma^{5} )d_{j}V_{ij}+\frac{g}{\sqrt{2}} W_{\mu}^{+}\overline{d}_{i} \gamma^{\mu} (1-\gamma^{5} )u_{j}V_{ij}^{\ast}
\end{equation}
where with the symbol $u_j$; we have indicated the up, charm and top quarks, while with $d_j$; the down, strange and bottom quarks. The unitary matrix V is the so-called Cabibbo, Kobayanshi and Maskawa matrix and is responsible for the mixing of the different families of quarks.
Under the CP transformation, obtained by the simultaneous application of the charge conjugation (C) and the parity (P), the operator L is transformed in the following way:
\begin{equation}
(CP)L(CP)^{-1}=\frac{g}{\sqrt{2}} W_{\mu}^{-}\overline{d_{i}} \gamma^{\mu} (1-\gamma^{5} )u_{j}V_{ij}+\frac{g}{\sqrt{2}} W_{\mu}^{+}\overline{u}_{i} \gamma^{\mu} (1-\gamma^{5} )d_{j}V_{ij}^{\ast}\,\,\,.
\end{equation}
From this equation it is clear that, in order to have invariance under CP, it is  necessary and sufficient that the matrix V is real. On the other side it is known that a unitary n x n matrix can be expressed as a function of $(n - 1)^2$ parameters, of which $\frac{n(n-1)}{2}$ are real and $\frac{(n-1)(n-2)}{2}$ are complex and, therefore, the Cabibbo, Kobayashi and Maskawa matrix can be parameterized as a function of three real parameters, plus a complex phase. This is not the place for a complete presentation of the origin of CP violation. We just stress that it is compatible and naturally present in the Standard Model. This of course does not mean that the standard Model origin of CP violation is the only one. Other contributions are still possible on the indirect side of CP violation. The importance of this hypothetical contribution coming from the gravity should be a strong function of the mixing time, not of the masses of the quarks, because the gravitational acceleration is independent of the masses. This circumstance points our attention back to Good's consideration of the gravitational contribution to CP violation.  

Chardin reformulated Good's argument in terms of relative potentials and showed that the gravitational field on the surface of the Earth is of the right intensity, i.e., of the required order of magnitude to cause CP~violation. In particular, the mixing time of the kaon,
\begin{equation}
\Delta \tau = 5.9 \times 10^{10} \,\, \mathrm{s} \,\, \simeq 6\tau_{k_{s}}
\end{equation}
is long enough for the gravitational field of the Earth to attract the matter and repel the antimatter components of the $\mathrm{K} $ meson to induce a separation,
\begin{equation}
\Delta \zeta = g \Delta \tau^{2} \,\,,
\end{equation}
between them inducing regeneration, thus providing a mechanism for indirect CP~violation. The amount of such a contribution should be roughly the same order of magnitude as the spatial separation divided by the Compton wavelength:
\begin{equation}
\label{eq:chi}
\begin{gathered}
\chi=\dfrac{\Delta \zeta}{\Delta L_{\mathrm{K}}}
=\dfrac{g \tau_{k_{l}}}{\dfrac{\hbar}{m_{\mathrm{K}}c}}
=\Omega\dfrac{g\dfrac{\pi^2\hbar^2}{\Delta m^2c^4}}{\dfrac{\hbar}{m_{\mathrm{K}}c}}\\ 
\chi=\Omega\dfrac{\pi^2\hbar gm_{\mathrm{K}}}{\Delta m^2c^3}=\Omega\times0.88\times10^{-3} \,\,,
\end{gathered}
\end{equation}
which happens to be the same order as the CP~violation parameter, $\varepsilon$, as measured on Earth's surface. 
If we calculate (\ref{eq:chi}) given for example, the gravitational strength on the Moon's surface we would expect the measured $\varepsilon$ to be $80\%$ smaller than 
the $\varepsilon$ measured on Earth's surface assuming gravitational repulsion beween matter and antimatter. 
As mentioned above, we expect only $16\%$ of the CP~violation induced by gravity in an experiment performed in the gravitational field on the Moon. Since
\begin{equation}
\label{eq:R}
R=\frac{\Gamma(\mathrm{K}_{\mathrm{L}}\rightarrow \pi^{+} \pi^{-} )}{\Gamma ((\mathrm{K}_{\mathrm{L}}\rightarrow \pi^{+} \pi^{-} \pi^{0} )}
\end{equation}
\noindent is quadratic in epsilon, we expect a very large difference in the value of $R$ when measured on the surface of the Moon.

\subsection{A CPV experiment in a low gravitational field, e.g., in a low Earth orbit}

 In several papers, our group has suggested an approach for performing a CPV experiment on the neutral kaon system in orbit \cite{piacentino-2016, piacentino-2017, piacentino-2019}. The idea is to use the flux of cosmic protons to generate the short- and long-lived kaons on a specialized target and to study the decay in presence of a gravitational field different from that on the Earth's surface. For details of the idea we refer the reader to our earlier papers~\cite{piacentino-2016, piacentino-2019}. At this moment we would like to underline that thanks to missions like Pamela and AMS-01/02 we now know very well the fluxes of cosmic protons in orbit measured by AMS-02~\cite{ams-02} and PAMELA~\cite{pamela-2017}. In the following, we considered the proton flux as determined by AMS-02.

\subsection{Feasibility study}

\begin{figure}[htp]
    \centering
    \includegraphics[trim={0cm 0cm 0cm 0cm}, clip, width=15.0cm, angle=0, scale=0.5]{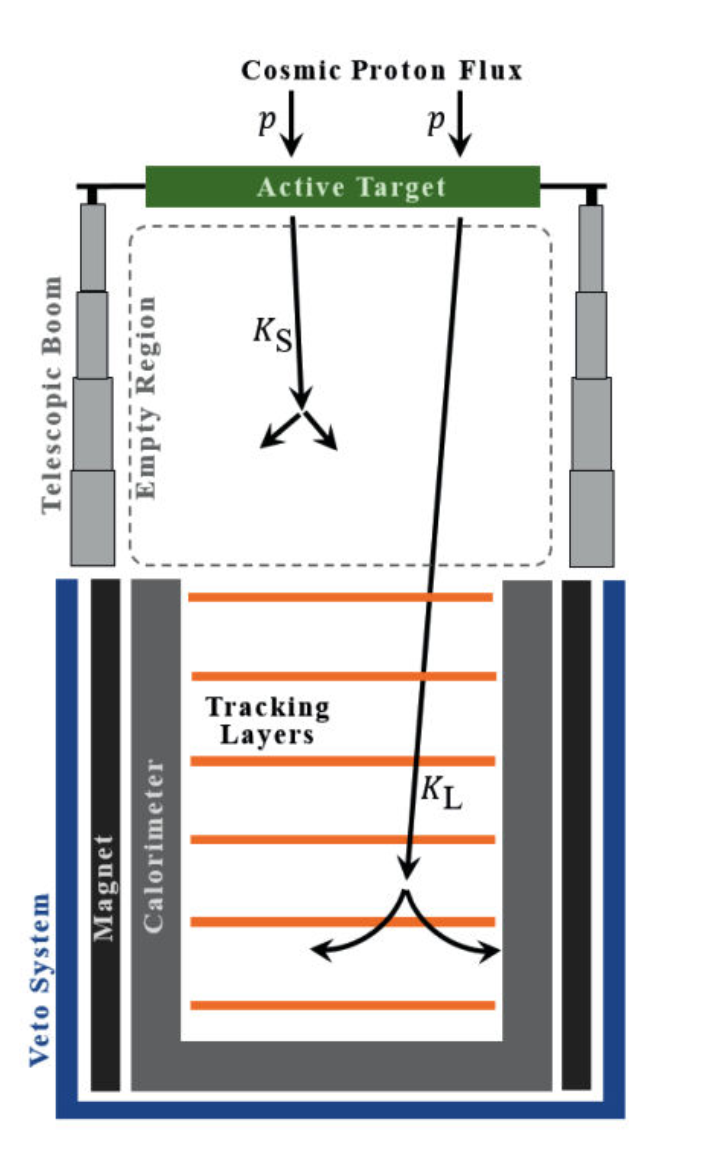}
    \caption{Notional structure of our proposed apparatus to measure the rate of CP-violating $\mathrm{K}_{\mathrm{L}}$ decays in a low-gravity environment.}
    \label{fig:apparatus}
\end{figure}

In our paper~\cite{piacentino-2016} we published a study in which we simulated the production of neutral kaons by cosmic protons hitting a target, using the cosmic proton energy spectrum measured by AMS-01 ~\cite{ams01_PLB427}. Our study investigated kaon production with a variety of target materials. The study also simulated the distance from the target at which the $\mathrm{K}_{\mathrm{L}}$ and $\mathrm{K}_{\mathrm{S}}$ decayed. In our subsequent work~\cite{piacentino-moriond-2021}, instead, we hypothesized a detector, to be positioned in low Earth orbit (LEO), designed to study the decay of neutral kaons that, produced by the impact of cosmic protons on a suitable target, ended up decaying in the sensitive volume of the detector. The structure of the apparatus is shown in Figure~\ref{fig:apparatus}.
The main components are:
\begin{enumerate}
\item an active Target capable of providing the time and position of the origin of each recorded event;
\item a veto system that excludes events not related to Kaon production in the Target;
\item a magnetic field produced with permanent magnets;
\item a tracking system;
\item an electromagnetic calorimeter; and
\item a time-of-flight (TOF) system, not shown in the figure.
\end{enumerate}

At that time, we did not yet study the acceptance of the apparatus, nor were we absolutely certain that this design would be able to identify the decays of neutral kaons in the right way to draw conclusions on CP violation as described in the previous paragraphs. The decays of interest to us are:
\begin {enumerate}
\item two-pion decay with opposite charge;
\item three-pion decay, two of opposite charge and one neutral;
\item decays containing muons, representing an important background for our proposed measurement.
\end{enumerate}

Fortunately, we can take advantage of the results of a very similar experiment that was performed during the AMS-01 mission aboard the space shuttle \emph{Discovery} that, in 1998, was in orbit while docked to the Russian Space Station MIR. On that occasion, AMS-01 performed a very interesting measurement by obtaining the image of a part of the space station, by collecting the decay daughters of the particles produced by the collisions between cosmic protons and the structures constituting the MIR Station. In other words, the space station acted as a target for cosmic protons, thereby producing secondary particles. Even the simple analysis of the publications on the subject represents a precedent of extreme interest for the evaluation of the feasibility of the measurement we propose. Extracting them from the same publication, we present crucial figures and data to demonstrate the scientific opportunity of further interesting investigations~\cite{AGUILAR2005321}.

\begin{figure}[htp]
    \centering
    \includegraphics[trim={0cm 0cm 0cm 0cm}, clip, width=15.0cm, angle=0, scale=0.7]{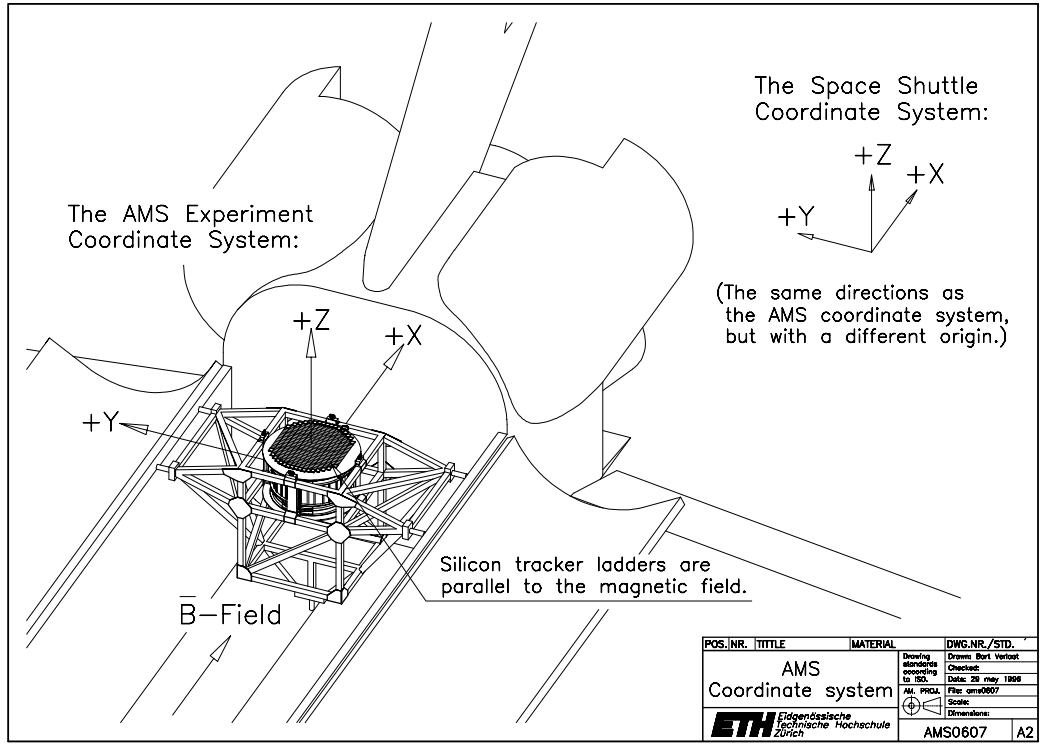}
    \caption{Diagram of the AMS-01 detector aboard the Shuttle Discovery (taken from~\cite{AGUILAR2005321}).}
    \label{fig:ams-shuttle}
\end{figure}

\begin{figure}[htp]
    \centering
    \includegraphics[trim={0cm 0cm 0cm 0cm}, clip, width=15.0cm, angle=0, scale=0.9]{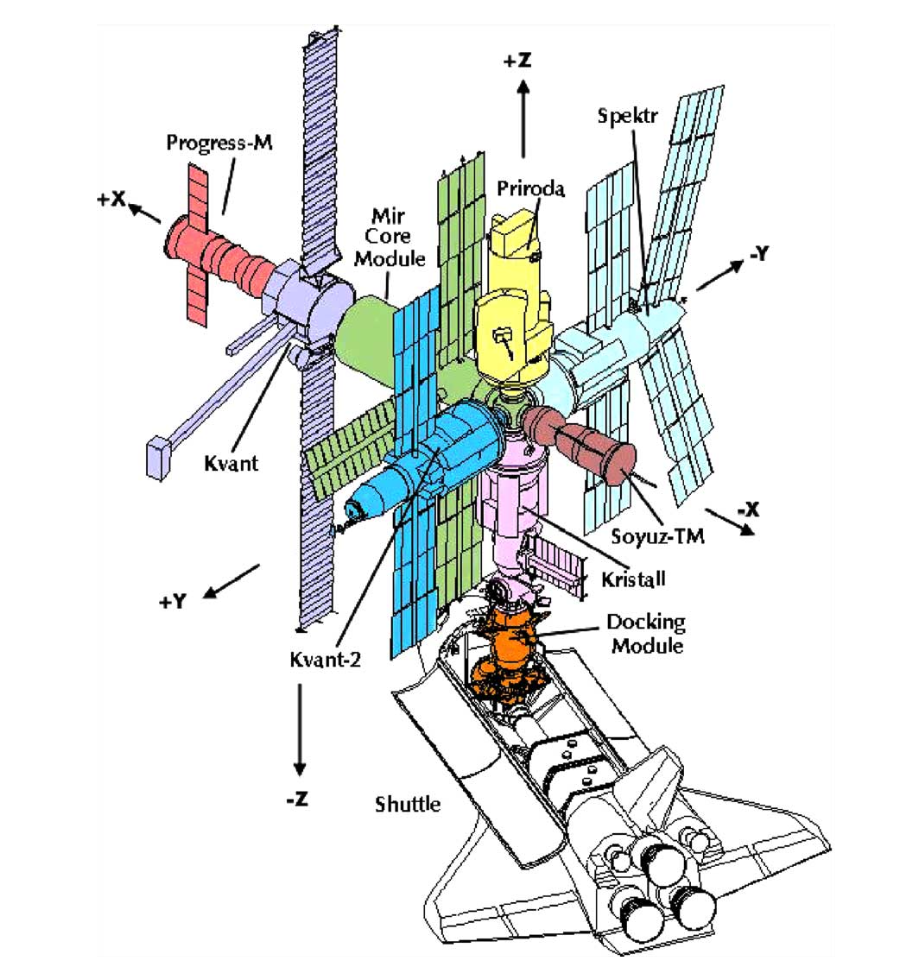}
    \caption{Illustration of Space Shuttle Discovery docked with the Mir space station (taken from~\cite{AGUILAR2005321}).}
    \label{fig:shuttle-docked-mir}
\end{figure}

Figure~\ref{fig:ams-shuttle} shows the position of AMS-01 inside the cargo bay of \emph{Discovery}, 
and Figure~\ref{fig:shuttle-docked-mir} shows the docking scheme that the space shuttle \emph{Discovery} used during the 1998 mission. In this second image, the modules of the MIR Station are visible. In particular are evident the three docked vehicles, namely:
\begin{itemize}
\item a Progress Vehicle for the transport of materials;
\item a Soyuz for the transport of people;
\item the Shuttle Discovery for the transport of people and things. 
\end{itemize}

As can be seen in Figure~\ref{fig:shuttle-docked-mir}, the Shuttle is in a position such that the cosmic rays intercepting the Space Station can give decay products detectable by AMS1. These decay products have actually been detected and published and our subsequent considerations refer to them.

\begin{figure}[htp]
    \centering
    \includegraphics[trim={0cm 0cm 0cm 0cm}, clip, width=15.0cm, angle=0, scale=0.5]{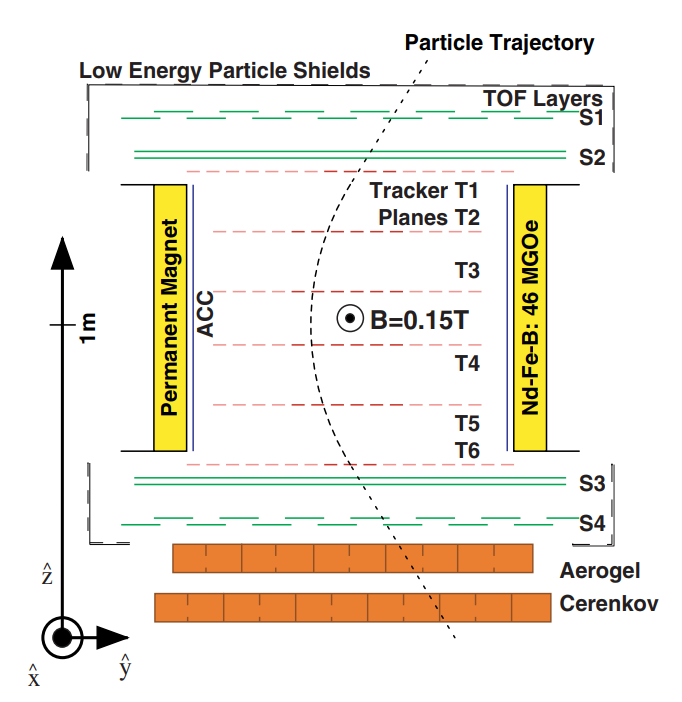}
    \caption{Structure of the AMS-01 detector (taken from~\cite{AGUILAR2005321}).}
    \label{fig:ams-structure}
\end{figure}

The structure of AMS-01, visible in Figure~\ref{fig:ams-structure}, is made up of a TOF system, a magnetic field of 0.15~Tesla generated inside a hollow cylindrical array of permanent Nd-Fe-B magnets, a Tracker made of silicon strips and an Aerogel Threshold Cerenkov detector for improved particle identification. The resolution of $\frac{\delta \beta}{\beta}$, was of the order of 3\% while the resolution of the rigidity $\frac{\delta R}{R}$ was in the range between 150 and 200~GeV/c with an optimal resolution of 7\% at about 10~Gev/c. Measuring $\frac{\delta E}{\delta x}$ in the TOF and in the silicon strips, the charge was traced. From the rigidity and the direction of curvature the sign and the moment were measured, finally from the TOF and the Cerenkov, the velocity of each particle was measured. It was thus possible to calculate the mass of the particle that passed through the detector and consequently identify it according to the formula:
\begin{equation}
m=\frac{\left| Z \right| R}{c} \sqrt{\beta^{-2} -1} 
\end{equation}
where $p=R$ for $Z=1$ particles.

\begin{figure}[htp]
    \centering
    \includegraphics[trim={0cm 0cm 0cm 0cm}, clip, width=15.0cm, angle=0]{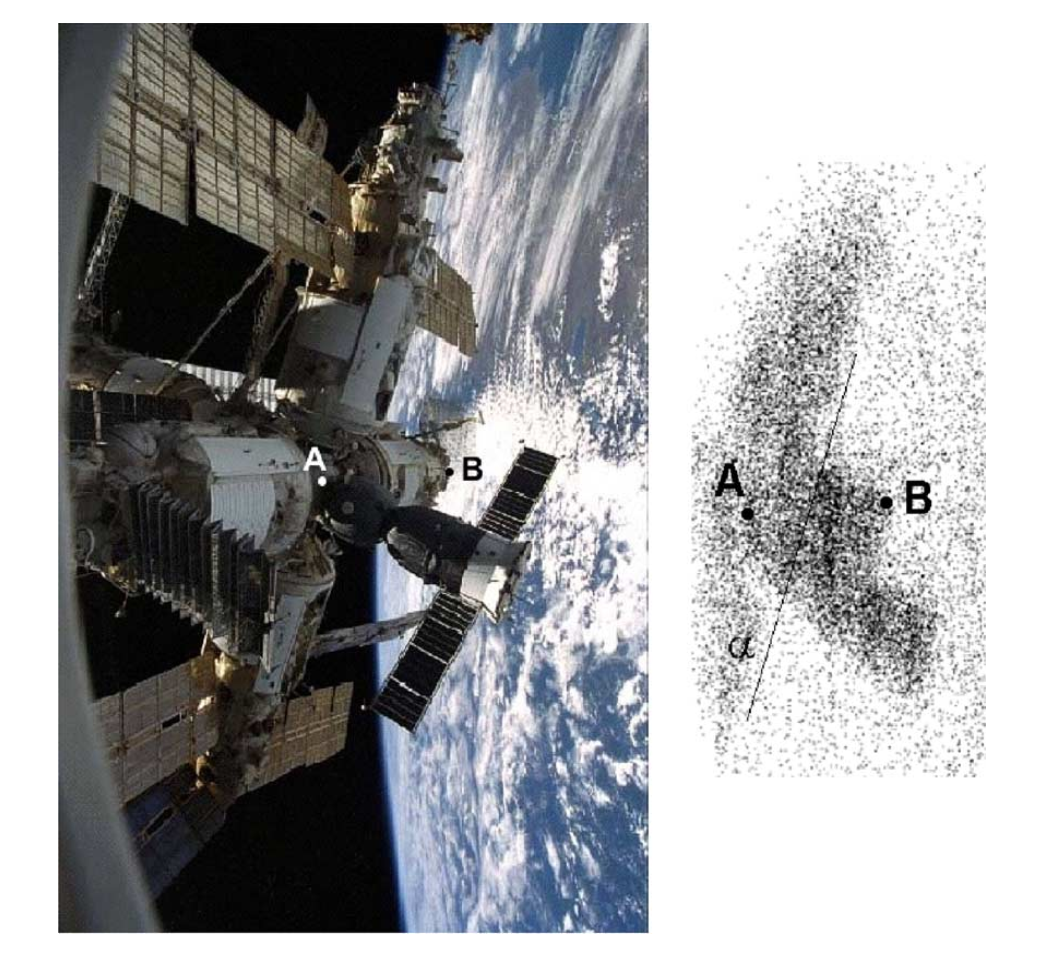}
    \caption{Photograph of the Mir Space Station (\emph{left}) and reconstruction of the Mir Space Station from daughter particles originating from cosmic protons interacting with the space station (images taken from~\cite{AGUILAR2005321}).}
    \label{fig:image-of-mir-from-daughters}
\end{figure}

Figure~\ref{fig:image-of-mir-from-daughters} (\emph{right}) shows the image created by AMS-01 on the basis of the secondary particles produced by the cosmic protons interacting with the MIR space station. If we compare this image with a photograph of the Mir Space Station, Figure~\ref{fig:image-of-mir-from-daughters} (\emph{left}), we can recognize the various modules, e.g., the Soyuz, and we verify that the solar panels are too thin to create their own image. In the original study, these considerations were made in an accurate and quantitative manner and here we refer to the original data table for their study~\cite{AGUILAR2005321}. 

\begin{figure}[htp]
    \centering
    \includegraphics[trim={0cm 0cm 0cm 0cm}, clip, width=15.0cm, angle=0, scale=0.5]{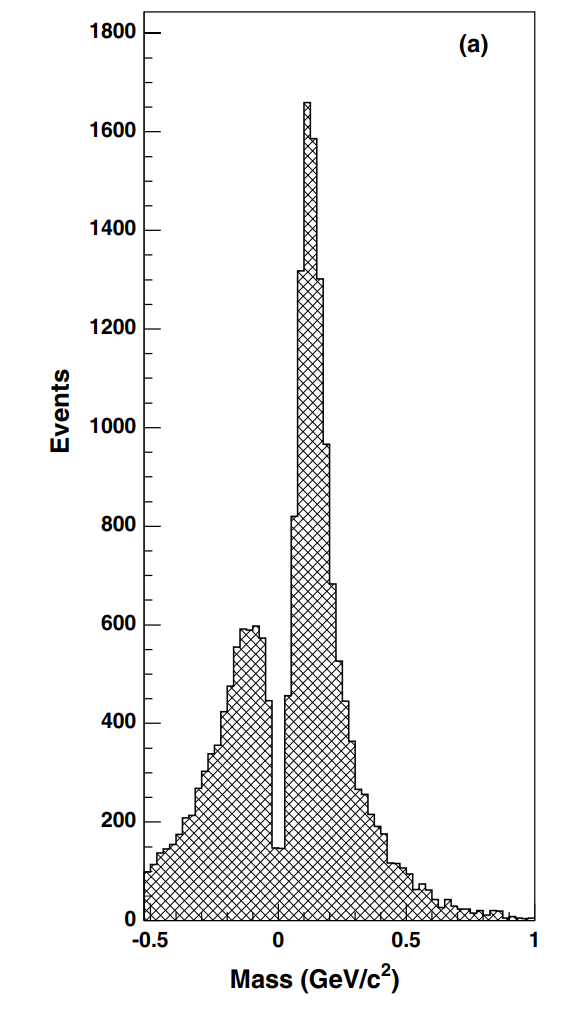}
    \caption{Mass distribution of particles originating in the Mir space station, demonstrating the ability to distinguish muons from pions (plot taken from ~\cite{AGUILAR2005321}).}
    \label{fig:ams-mass-distribution}
\end{figure}

In Figure~\ref{fig:ams-mass-distribution}, we include a plot of the mass distribution that they obtained. If we consider the mass spectrum of the negative unit charge particles recorded during the data taking of the Mir image by AMS-01, we note a high mass peak in the area between 0.1 and 02 GeV/c$^{2}$ and therefore they can be identified with certainty as a muon or a pion. Further, their spectrum at the moment suggests that the pions come from no more than ten meters. That is, they are decay products originating from the interaction between cosmic protons and the space station. The cited article also calculates the equivalent thickness of the Mir Space Station with respect to the cosmic flux, estimating it at about ten centimeters of aluminum. This consideration is particularly useful for the sizing of a target for the experiment we intend to propose.

\begin{figure}[htp]
    \centering
    \includegraphics[trim={0cm 0cm 0cm 0cm}, clip, width=15.0cm, angle=0, scale=0.5]{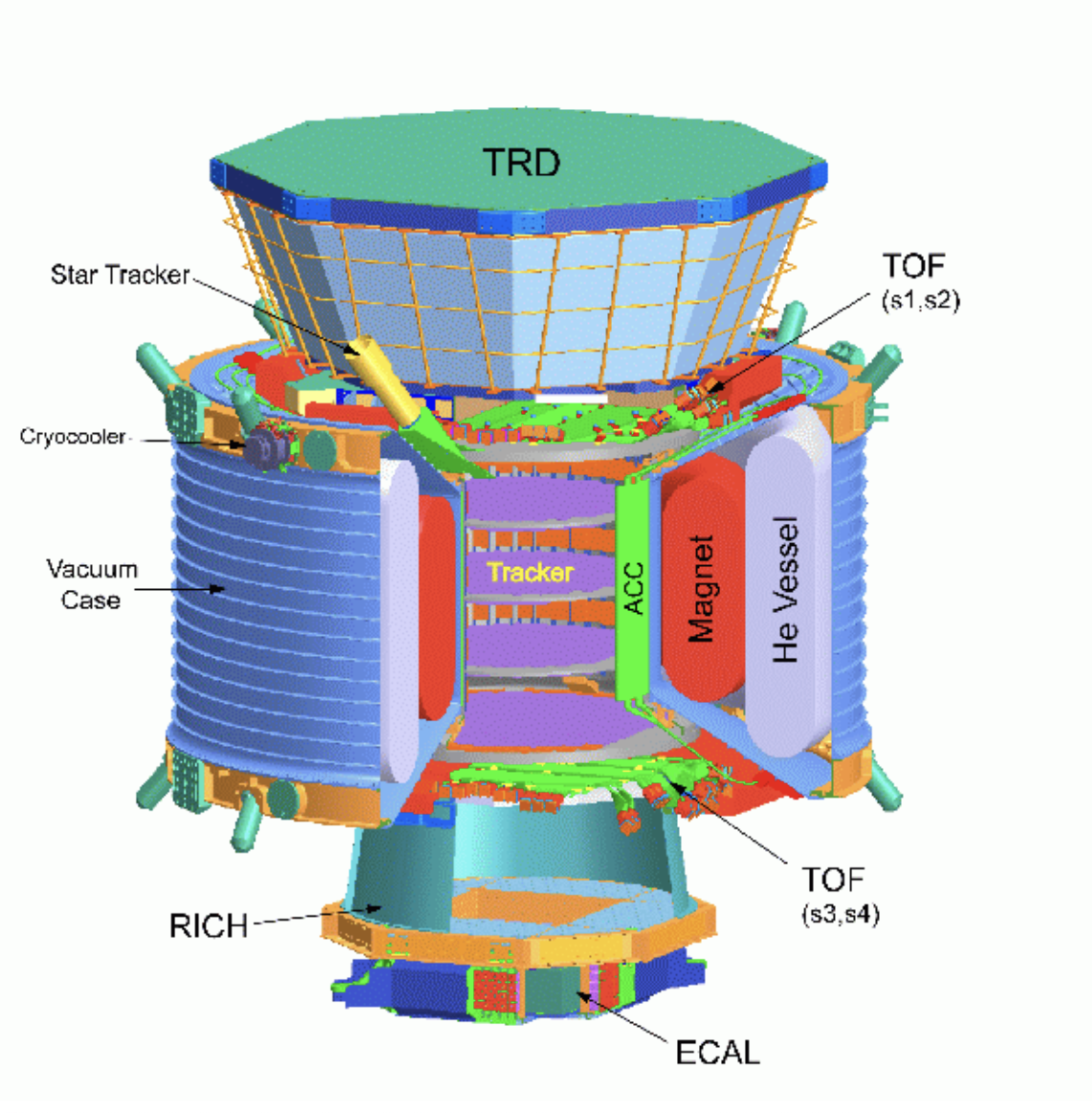}
    \caption{Structure of the AMS-02 detector (taken from~\cite{ARRUDA200732}).}
    \label{fig:ams-02-apparatus}
\end{figure}

Figure~\ref{fig:ams-02-apparatus} shows the structure of AMS-02, which is still collecting data on the ISS. It can be seen that this detector comprises all but one of the elements that appear in the structure of the detector we proposed in one of our previous works~\cite{piacentino-moriond-2021}. Indeed, it contains the TRD detector that could be used as a detector of the decay vertex of neutral kaons; however, the presence of a dedicated target is missing. Here, we suggest that within the activities of AMS-02, a Monte Carlo study could be carried out, hypothesizing a target placed at a certain distance away, along the central axis of the AMS-02 detector. The parameters of thickness, distance, and material of such a target could be investigated for the purpose of carrying out our proposed measurement. The possible execution of preliminary measurements, or even the realization of the experiment, would constitute the first study of fundamental interactions as a function of the gravitational field.

\section{Conclusion}

In this paper, we point out a potential difference in the direction of the gravitational push or pull when considering composite and non-composite antimatter. We reviewed three theoretical frameworks that permit, or predict, matter-antiparticle gravitational repulsion. In doing so, we also show that the ALPHA-$g$ experiment's recent results---while conclusively demonstrating that antihydrogen falls toward the Earth---may actually indicate a repulsive component at the level of $\sim$15\% due to the antiparticles themselves. Additionally, we revisit our previously proposed low-gravity neutral kaon CP-violation experiment. We suggest that confidence in our proposed experiment's feasibility may be increaesd by performing Monte Carlo studies and data analysis on existing AMS-01 data, specifically on the reconstruction of pion and muon decay daughters from cosmic protons interacting with the Mir space station. 

\section{Acknowledgments}

We would like to thank the Rencontres de Moriond Conference on Gravitation reviewers for useful feedback and for including our work in the conference. Additionally, we would like to thank Barbara Pasquini and Claudio Corianò for fruitful discussions. In particular, we would also like to thank Graziano Venanzoni, both for the encouragement and for the important discussions, but even more for his notable contributions to the initial phases of this study.

\bibliographystyle{plain} 
\bibliography{references} 

\end{document}